\documentclass[pra,aps,twocolumn,superscriptaddress,longbibliography,floatfix,notitlepage]{revtex4-1}

\usepackage{graphicx} 
\usepackage{amsmath} 
\usepackage{multirow}
\usepackage{float}
\usepackage{hyperref}
\usepackage{xcolor,tikz, circuitikz}
\usepackage{stmaryrd}

\hypersetup{
colorlinks,
linkcolor={blue},
citecolor={blue},
urlcolor={blue!80!black}
}

\usepackage{algpseudocode}
\usepackage{newfloat}

\DeclareFloatingEnvironment[
    fileext=loa,
    listname=List of Algorithms,
    name=ALGORITHM,
]{algorithm}

\AtEndEnvironment{algorithm*}{\noindent\hrulefill\par\nobreak\vskip-5pt}

\newcommand{\ket}[1]{|{#1}\rangle}
\newcommand{\bra}[1]{\langle{#1}|}
\newcommand{\ketbra}[1]{|{#1}\rangle\langle{#1}|}
\newcommand{\kb}[2]{|{#1}\rangle\langle{#2}|}

\begin{document}

\title{\LARGE \bf
Resource Reduction in Multiparty Quantum Secret Sharing of both Classical and Quantum Information under Noisy Scenario
}

\author{Nirupam Basak}
\email{nirupambasak2020@iitkalumni.org}
\affiliation{Cryptology and Security Research Unit, Indian Statistical Institute, Kolkata 700108, India}

\author{Goutam Paul}
\email{goutam.paul@isical.ac.in}
\affiliation{Cryptology and Security Research Unit, Indian Statistical Institute, Kolkata 700108, India}

\date{\today}

\begin{abstract}
Quantum secret sharing (QSS) enables secure distribution of information among multiple parties but remains vulnerable to noise. We analyze the effects of bit-flip, phase-flip, and amplitude damping noise on the multiparty QSS for classical message (QSSCM) and secret sharing of quantum information (SSQI) protocols proposed by Zhang et al. (Phys. Rev. A, 71:044301, 2005). To scale down these effects, we introduce an efficient quantum error correction (QEC) scheme based on a simplified version of Shor’s code. Leveraging the specific structure of the QSS protocols, we reduce the qubit overhead from the standard 9 of Shor's code to as few as 3 while still achieving lower average error rates than existing QEC methods. Thus, our approach can also be adopted for other single-qubit-based quantum protocols. Simulations demonstrate that our approach significantly enhances the protocols’ resilience, improving their practicality for real-world deployment.
\end{abstract}

\maketitle

\section{INTRODUCTION}

Consider a national government operating a secure server that hosts highly sensitive data vital to national security. Any mishandling or unauthorized access to this information could lead to catastrophic consequences, underscoring the risk of entrusting the server's passkey to a single individual. This prompts a fundamental question: \emph{how can the passkey be stored securely?} A promising solution is to divide the passkey among multiple trusted personnel, such that only a designated subset can collaboratively reconstruct and access it. Implementing this approach necessitates robust protocols for both secure distribution and reliable reconstruction of the key. Secret sharing protocols~\cite{10.1145/359168.359176,8817296,CHATTOPADHYAY2024100608,panagopoulos2010secretsharingschemeusing,cryptoeprint:2012/246,mashhadi2016fairly} offer a well-established framework to address this challenge.


Traditional secret sharing schemes derive their security from the computational hardness of certain mathematical problems, including polynomial interpolation, integer factorization, and discrete logarithms~\cite{panagopoulos2010secretsharingschemeusing, cryptoeprint:2012/246,doi:10.1049/el:19941076}. However, the advent of quantum computing poses a significant threat to these classical foundations, as quantum algorithms are capable of efficiently solving problems that underpin the security of these schemes~\cite{365700,10.1145/237814.237866}.

Quantum Secret Sharing (QSS)\cite{PhysRevA.59.1829,PhysRevA.71.044301,PhysRevA.69.052307,GUO2003247,PhysRevA.61.042311,PhysRevLett.121.150502,PhysRevA.103.032410,PhysRevA.63.042301}, by contrast, harnesses the principles of quantum mechanics—such as superposition and entanglement—to distribute secrets in a fundamentally different manner. Instead of depending on the computational hardness of mathematical problems, QSS leverages intrinsic quantum properties\cite{heisenberg1927anschaulichen,wootters1982single} to provide enhanced security. This makes QSS particularly compelling in the emerging quantum era, where conventional cryptographic methods may be rendered obsolete by powerful quantum algorithms.

In a QSS protocol, the secret is encoded into multiple quantum bits (qubits) and distributed among participants. Due to the nature of quantum mechanics, a participant holding only one share cannot extract any meaningful information about the secret without disturbing the quantum system, making unauthorized observation or interception detectable~\cite{wootters1982single,basak2023quantum}. To recover the original secret, a predefined minimum number of participants must collaborate and perform coordinated quantum operations on their respective shares. A QSS scheme that distributes the secret among $n$ parties and requires at least $k$ of them to reconstruct it is referred to as an $(n,k)$-QSS scheme.


Quantum noise~\cite{Preskill2018quantumcomputingin,nielsen2010quantum,boixo2018characterizing} poses a major challenge to QSS, as it can disturb the fragile quantum states that the system relies on~\cite{nielsen2010quantum,gottesman2009introductionquantumerrorcorrection,barends2014superconducting}. Arising from the fundamental principles of quantum mechanics, such as uncertainty and decoherence, quantum noise introduces fluctuations that affect the transmission and measurement of quantum information. In the context of QSS, such noise can degrade the quality of the distributed qubit shares, making it difficult for participants to retrieve a sufficient number of intact shares for successful secret reconstruction. If too many shares are corrupted, the protocol may fail to meet the threshold required, rendering the secret unrecoverable~\cite{basak2023quantum}.


A key advantage of QSS is its inherent ability to detect eavesdropping: any attempt by an unauthorized party to intercept or measure the quantum shares typically introduces detectable disturbances~\cite{joy2020implementation,basak2023quantum}. However, excessive quantum noise can obscure these disturbances, making it difficult to distinguish between natural errors and deliberate interference~\cite{basak2023quantum}. This compromises one of the core security features of QSS and highlights the need for robust noise mitigation strategies.

To mitigate the effects of quantum noise, a range of strategies has been developed~\cite{Roffe03072019,RevModPhys.95.045005,PhysRevLett.81.2152,devitt2013quantum}, among which quantum error correction (QEC) techniques play a central role~\cite{PhysRevLett.81.2152,PhysRevA.52.R2493,PhysRevLett.79.3306,doi:10.1142/S0217984997001304,PhysRevA.73.060302,PhysRevA.54.4741,PhysRevLett.77.793,PhysRevA.54.1862,PhysRevA.57.127}. QEC allows a quantum system to detect and correct errors introduced by noise, thereby preserving the integrity of the quantum information. In the context of QSS, these techniques are essential for ensuring that the distributed qubit shares remain usable. By integrating QEC, QSS protocols become significantly more resilient to noise, enabling authorized participants to reliably reconstruct the secret even in the presence of environmental disturbances.

The fundamental principle of QEC is to encode quantum information in a way that distributes it across multiple physical qubits. This redundancy enables the system to detect and correct errors affecting individual qubits without compromising the encoded information. For instance, a single logical qubit, representing a unit of quantum information, can be encoded into a group of physical qubits. If one of these qubits is altered by noise, the error can be identified through specific measurements on the others, allowing the original quantum state to be accurately restored.

One of the most notable quantum error correction codes is Shor's code~\cite{PhysRevA.52.R2493}, which was the first to demonstrate that quantum information could be safeguarded from errors by using ancillary qubits. Shor's code encodes a single logical qubit into nine physical qubits, allowing it to correct arbitrary errors affecting any one of these qubits. 

\noindent{\em Our Contributions.} In this article, we explore multiparty Quantum Secret Sharing of classical messages (QSSCM) and secret sharing of quantum information (SSQI) protocols~\cite{PhysRevA.71.044301} proposed by Zhang et al., which utilize single-qubit states. These protocols are straightforward to implement, as they do not involve entanglement generation or multi-qubit quantum operations, except for the teleportation step in the SSQI protocol. In these protocols, each participant performs simple Pauli or Hadamard operations before forwarding the qubit to the next party. However, the transmission channel between parties may introduce noise, potentially corrupting the quantum state. To mitigate this, we employ Shor's 9-qubit code for error protection. By exploiting the specific structure of the protocol, we demonstrate that certain parts of the code can be bypassed. In particular, we only require the bit-flip and phase-flip error correction codes, reducing the qubit overhead from 9 to 3. In general. such $3$-qubit abridged version of Shor's code does not correct amplitude damping noise. However, here it works due to the structure of the QSS protocol. This modified 3-qubit code can also be used for other single-qubit-based QSS~\cite{GUO2003247,refId0,PhysRevA.92.030301}, quantum key distribution (QKD)~\cite{1573668924803137792,PhysRevLett.68.3121, PhysRevA.76.062316, PhysRevLett.92.057901}, quantum secure direct communication (QSDC)~\cite{PhysRevA.69.052319,WANG2006256,10.1007/s11128-017-1757-x,JiXin1418} and quantum authentication (QA)~\cite{10.1007/s11128-014-0767-1,4099193,doi:10.1142/S0219749916500027,PhysRevA.62.022305,10.1007/s11128-018-2124-2} protocols. Our results show that this modified code effectively minimizes errors in the reconstructed secret. Moreover, this modified code performs better than the existing QEC codes.

\noindent{\em Paper Outline.} In Section~\ref{sec:QSSCM}, we briefly revisit the multiparty QSSCM protocol provided by Zhang et al. Then, we discuss the quantum noise models and the QEC codes we considered in Section~\ref{sec:noise_QEC}. The effect of noise on the above QSSCM protocol has been discussed in Section~\ref{sec:noise}. The reduction of error after using QEC is shown in Section~\ref{sec:QEC}. In Section~\ref{sec:SSQI}, we discuss the SSQI protocol under noise and effect of QEC. Finally, in Section~\ref{sec:conclusion}, we conclude our work.

\section{\label{sec:QSSCM}Revisiting Multiparty QSSCM}

The QSSCM protocol~\cite{PhysRevA.71.044301} designed by Zhang et al. is based on a previously developed protocol for quantum secure direct communication by  Deng and Long~\cite{PhysRevA.69.052319}. In this scheme, a sender, Alice, splits her secret into encrypted shares and distributes them to different receivers. Each receiver applies certain quantum operations to ensure security before passing the message along. The receivers can only recover the full message by working together, ensuring that no individual can access it alone. Thus, it is an $(n,n)$-QSS protocol.

\begin{algorithm*}[htpb]
\hrulefill
\vskip-10pt
\caption{Multiparty QSSCM Scheme}\label{algo:QSSCM}
\vskip-6pt\hrulefill
\begin{algorithmic}[1]
\State \textbf{Sender: }Alice, \textbf{Receiver: }Bob, Charlie, Dave,\dots, Zach.
\State\label{step:bob} Bob prepares a batch of single qubits $\{\ket{\psi_i}\}_i$ randomly from $\{\ket{0}, \ket{1}, \ket{+}, \ket{-}\}$ and sends the qubits to the next receiver, Charlie.
\begin{equation*}
\ket{\psi_i}=U_B\ket{0}\in\{\ket{0}, \ket{1}, \ket{+}, \ket{-}\}.
\end{equation*}
\State\label{step:charl} After receiving these qubits, for each qubit, Charlie randomly chooses a unitary operator $U_C$ from $\{I, \sigma_y, H\}$ and applies this operator to the qubit. Here, $I$ is the identity operator $\ketbra{0}+\ketbra{1}$, $\sigma_y$ is Pauli $Y$ operator $\ket{0}\bra{1}-\ket{1}\bra{0}$ and $H$ is the Hadamard operator $\frac{1}{\sqrt{2}}[\ketbra{0}+\ket{0}\bra{1}+\ket{1}\bra{0}-\ketbra{1}]$. After this encryption, she sends the batch to Dave.
\begin{equation*}
\ket{\psi_i}\to U_C\ket{\psi_i},\;U_C\in\{I, \sigma_y, H\}.
\end{equation*}
\State\label{step:oth_rec} Dave randomly encrypts the encoded photons using the same method as Charlie, then forwards them to the next receiver. Each participant repeats this process until Zach completes his encryption. Once finished, Zach sends the fully encrypted photons to Alice.
\begin{equation*}
U_C\ket{\psi_i}\to U_Z\cdots U_DU_C\ket{\psi_i},\;U_Z,\cdots,U_D\in\{I, \sigma_y, H\}.
\end{equation*}
\State\label{step:alic} Alice performs some security check. Upon success, she discards the qubits used in security checking, and encodes her secret by applying unitary $U_A$, which is either $I$ (for $0$) or $\sigma_y$ (for $1$), on the remaining qubits. Finally, she forwards these qubits to Charlie.
\begin{equation*}
U_D\cdots U_C\ket{\psi_i}\to U_AU_D\cdots U_C\ket{\psi_i},\;U_A\in\{I, \sigma_y\}.
\end{equation*}
\State\label{step:recon} If Bob and Charlie collaborate, they can reconstruct the secret. First, they apply the inverse of their respective operations in step~\ref{step:bob},~\ref{step:charl} and~\ref{step:oth_rec}. Then they measure the state in computational basis $\{\ket{0}, \ket{1}\}$ to get the secret.
\begin{equation*}
U_AU_D\cdots U_C\ket{\psi_i}\to U_B^\dagger U_C^\dagger \cdots U^\dagger_ZU_AU_Z\cdots U_CU_B\ket{0}=U_A\ket{0}.
\end{equation*}
\end{algorithmic}
\end{algorithm*}

The multiparty QSSCM protocol~\cite{PhysRevA.71.044301} is provided as Algorithm~\ref{algo:QSSCM}. Note that the states produced by Bob in step~\ref{step:bob}, are the basis elements of the computational basis and the Hadamard basis. Also, all the operations the participants apply are either commutative or anti-commutative, producing a global phase $\pm1$. Therefore, ignoring the global phase, the ordering of the operations may be changed in step~\ref{step:recon} to get the Alice's secret. Thus, after applying the operations, the state of the qubits would become $U_A\ket{0}$ up to global phase $\pm1$, where $U_A$ is the operation applied by Alice. Now, by measuring these qubits on a computational basis, Bob and Charlie can get the operation applied by Alice, revealing Alice's secret. Mathematically, the protocol grows as follows.
\begin{align}
&U_B^\dagger U_C^\dagger \cdots U^\dagger_ZU_AU_Z\cdots U_CU_B\ket{0}\notag\\
=&\pm U_B^\dagger U^\dagger_C\cdots U_AU_Z^\dagger U_Z\cdots U_CU_B\ket{0}\notag\\
=&\cdots=\pm U_B^\dagger U_AU_B\ket{0}\notag\\
=&\pm U_AU_B^\dagger U_B\ket{0}=\pm U_A\ket{0},
\end{align}
where, $U_A, U_B$ and $U_C$ are the unitary operations by Alice, Bob and Charlie, respectively.

Although the protocol works perfectly in the ideal scenario, the noise in the communication channels corrupts the qubits and makes it hard for the receivers to reconstruct the shared secret. In Section~\ref{sec:noise} and ~\ref{sec:QEC}, we discuss the effect of channel noises on the protocol and how to reduce the error in the reconstructed secret using QEC.



\section{\label{sec:noise_QEC}Quantum Noise and Error Correction}

Quantum noise makes any quantum protocol hard to implement~\cite{nielsen2010quantum}. It changes the state of a qubit and leads to an erroneous result at the end. Several QEC codes~\cite{PhysRevA.52.R2493,laflamme1996perfect,bennett1996mixed,leung1997approximate,dutta2025smallestquantumcodesamplitude,doi:10.1098/rspa.1996.0136,PhysRevLett.77.793,basak2025approximatedynamicalquantumerrorcorrecting,PhysRevA.75.012338,crepeau2005approximatequantumerrorcorrectingcodes,Hastings2021dynamically,PRXQuantum.4.020341,Haah2022boundarieshoneycomb} to protect quantum information from noise.

\subsection{\label{ssec:noise_model}Noise Models}

There are several noise models~\cite{PhysRevLett.120.050505,PhysRevLett.124.130501,nielsen2010quantum,basak2023quantum} available for quantum channels. However, as Pauli $X$ ($\sigma_x=\ket{0}\bra{1}+\ket{1}\bra{0}$) and Pauli $Z$ ($\sigma_z=\ketbra{0}-\ketbra{1}$), along with $I$ and $i\sigma_y=i\sigma_z\sigma_x$, form a basis for single qubit states, most of the single qubit noises, including depolarizing noise, can be easily transformed into a combination of bit-flip and phase-flip noise~\cite{nielsen2010quantum}. So, in this work, we are going to consider three common noises: bit-flip, phase-flip and amplitude damping.

\paragraph{\label{para:bit_flip_noise}Bit-flip Noise}

A bit-flip noise flips a state in the computational basis, i.e., it interchanges $\ket{0}$ and $\ket{1}$ up to some probability, called \emph{bit-flip error probability}. The Kraus operators~\cite{tong2006kraus} of a bit-flip channel $\mathcal{C}_b$ with bit-flip error probability $p_b^\mathcal{C}$ is given by $\left\{\sqrt{1-p_b^\mathcal{C}}I, \sqrt{p_b^\mathcal{C}}\sigma_x\right\}$. The action of the channel on some density matrix $\rho$ is as follows
\begin{equation}
\mathcal{C}_b(\rho)=(1-p^\mathcal{C}_b)\rho+p^\mathcal{C}_b\sigma_x\rho\sigma_x.
\end{equation}

\paragraph{\label{para:phase_flip_noise}Phase-flip Noise}

A phase-flip noise flips the relative phase of a state in computational basis up to some probability, called \emph{phase-flip error probability}. This error interchanges the states $\ket{+}$ and $\ket{-}$. The Kraus operators of a phase-flip channel $\mathcal{C}_p$ with phase-flip error probability $p_p^\mathcal{C}$ is given by $\left\{\sqrt{1-p_p^\mathcal{C}}I, \sqrt{p_p^\mathcal{C}}\sigma_z\right\}$ and the corresponding channel action on some density matrix $\rho$ is as follows
\begin{equation}
\mathcal{C}_b(\rho)=(1-p^\mathcal{C}_b)\rho+p^\mathcal{C}_b\sigma_z\rho\sigma_z.
\end{equation}

\paragraph{\label{para:amp_damp_noise}Amplitude Damping Noise}

Amplitude damping noise represents energy loss in a quantum system. It models the process where a qubit interacts with its environment and loses energy. This type of noise is particularly relevant in systems like superconducting qubits and optical quantum communication, where energy dissipation is a major concern~\cite{Chirolli01052008, PhysRevResearch.4.023034}. Mathematically, the action of an amplitude damping channel $\mathcal{C}_a$ with damping strength $\gamma\in[0,1]$ on some density matrix $\rho$ can be written as
\begin{equation}
\mathcal{C}_a(\rho)=E_0\rho E_0^\dagger+E_1\rho E_1^\dagger,
\end{equation}
where the Kraus operators are given by
\begin{equation}
E_0=\begin{pmatrix}
1&0\\
0&\sqrt{1-\gamma}
\end{pmatrix},\;
E_1=\begin{pmatrix}
0&\sqrt{\gamma}\\
0&0
\end{pmatrix}.
\end{equation}

Amplitude damping noise is crucial in quantum error correction and fault-tolerant quantum computing as it represents a primary source of decoherence in real-world quantum devices~\cite{Chirolli01052008}.

\subsection{\label{ssec:QEC}Error-correcting Codes}

For perfect protection of a qubit from an arbitrary noise, we require at least five qubits, due to quantum singleton bound~\cite{DJORDJEVIC2012227}. Although, some four- and three-qubit codes have been proposed~\cite{leung1997approximate,dutta2025smallestquantumcodesamplitude} to protect a qubit from amplitude damping noise, they are classified as approximate code, where error correction happens up to some threshold. Here we use the Shor's code~\cite{365700}, five-qubit perfect code~\cite{laflamme1996perfect} and four-qubit approximate code~\cite{leung1997approximate}. By exploiting the structure of the QSSCM protocol, we show that we can use the bit-flip and phase-flip codes separately to get 3-qubit repetition code for perfect protection of single qubit error. We also show that this 3-qubit repetition code outperforms the 4-qubit approximate code and the smallest perfect code of 5 qubits.

\paragraph{\label{para:Shor}Shor's Code}

Shor's code~\cite{PhysRevA.52.R2493}, introduced by Peter Shor in 1995, was one of the first quantum error correction codes. It protects a single logical qubit from arbitrary errors (bit-flip, phase-flip, and combinations of both) by encoding it into nine physical qubits. The code is composed of two parts as follows.

\textbf{Bit-flip error correction}: The first step of Shor's code involves encoding the logical qubit into three physical qubits. These qubits are encoded using a repetition code, where each qubit is copied three times to correct for bit-flip errors.

\textbf{Phase-flip error correction}: After the bit-flip error correction, the next step involves encoding each of the three qubits into another set of three physical qubits, using a three-qubit phase-flip code, which consists of a layer of Hadamard operation for change of basis, then copying the states three times, and finally another layer of Hadamard operation to go back to the original basis (or, in other words, copying each qubit three times in Hadamard basis). This helps to protect the information from phase-flip errors.

Together, these two layers of encoding (bit-flip and phase-flip corrections) allow Shor's code to protect a logical qubit from errors in both the bit and phase, as well as combinations of both, making it more robust against noise.

\paragraph{\label{para:5_qubit_code}Five-qubit Perfect Code}

Five-qubit code~\cite{laflamme1996perfect} is the smallest QEC code for perfect error correction. As the name suggests, the five-qubit code~\cite{PhysRevLett.77.793} encodes a single logical qubit using five physical qubits. The encoded logical qubits are as follows.
\begin{widetext}
\begin{equation}
\label{eq:QEC_encoding}
\begin{aligned}
\ket{0_L}&=\frac{1}{2\sqrt{2}}(-\ket{00000}+\ket{00110}+\ket{01001}+\ket{01111}-\ket{10011}+\ket{10101}+\ket{11010}+\ket{11100}),\\
\ket{1_L}&=\frac{1}{2\sqrt{2}}(-\ket{11111}+\ket{11001}+\ket{10110}+\ket{10000}+\ket{01100}-\ket{01010}-\ket{00101}-\ket{00011}).
\end{aligned}
\end{equation}
\end{widetext}

For QSSCM protocol, Bob performs this encoding operation on his qubits and sends the encoded states to Charlie. Charlie applies random logical operators from $\{I_L,\sigma_{yL}, H_L\}$ corresponding to the physical operators $\{I,\sigma_y,H\}$. Then Charlie sends the sequence to the next party. Finally, after performing security checks, Alice applies logical identity $I_L$ for secret bit $0$ and logical Pauli-$Y$ $\sigma_{yL}$ for secret bit $1$ and sends the sequence to Charlie. Upon receiving the sequence, all the receivers apply the logical operations they applied before Alice's encoding.

At the end, the receivers perform the decoding, which is simply the inverse of the initial encoding~\eqref{eq:QEC_encoding}, followed by  the state recovery operation, whose Kraus operators $\{\mathcal{R}_k\}_k$~\cite{PhysRevResearch.5.043161} are given by
\begin{widetext}
\allowdisplaybreaks
\begin{align*}
R_0&=\kb{00}{00}\otimes\sigma_0\otimes\kb{00}{00},&R_1&=\kb{00}{00}\otimes\sigma_z\otimes\kb{00}{01},\\
R_2&=\kb{00}{00}\otimes\sigma_0\otimes\kb{00}{10},&R_3&=\kb{00}{00}\otimes\sigma_0\otimes\kb{00}{11},\\
R_4&=\kb{00}{01}\otimes\sigma_0\otimes\kb{00}{00},&R_5&=\kb{00}{01}\otimes\sigma_z\otimes\kb{00}{01},\\
R_6&=\kb{00}{01}\otimes\sigma_x\otimes\kb{00}{10},&R_7&=\kb{00}{01}\otimes\sigma_x\otimes\kb{00}{11},\\
R_8&=\kb{00}{10}\otimes\sigma_0\otimes\kb{00}{00},&R_9&=\kb{00}{10}\otimes\sigma_x\otimes\kb{00}{01},\\
R_{10}&=\kb{00}{10}\otimes\sigma_z\otimes\kb{00}{10},&R_{11}&=\kb{00}{10}\otimes\sigma_x\otimes\kb{00}{11},\\
R_{12}&=\kb{00}{11}\otimes\sigma_z\otimes\kb{00}{00},&R_{13}&=\kb{00}{11}\otimes\sigma_x\sigma_z\otimes\kb{00}{01},\\
R_{14}&=\kb{00}{11}\otimes\sigma_x\otimes\kb{00}{10},&R_{15}&=\kb{00}{11}\otimes\sigma_z\otimes\kb{00}{11},\\
\end{align*}
\end{widetext}
where $\sigma_0$ is the identity operator and $\sigma_x,\sigma_z$ are the Pauli-$X$ and Pauli-$Z$ operators, respectively. After discarding the ancillary qubits and measuring the main qubit in the computational basis, they get the secret shared by Alice.

\paragraph{\label{para:4_qubit_code}Four-qubit Approximate Code}

The four-qubit approximate quantum error-correcting code~\cite{leung1997approximate} introduced by Leung et al. is designed to protect against amplitude damping errors, which commonly occur in realistic quantum systems due to energy loss. It encodes a single logical qubit into four physical qubits as
\begin{equation}
\label{eq:four_qubit_encode}
\begin{aligned}
|0_L\rangle&=\frac{1}{\sqrt{2}}(|0000\rangle+|1111\rangle),\\
|1_L\rangle&=\frac{1}{\sqrt{2}}(|0011\rangle+|1100\rangle).
\end{aligned}
\end{equation}
Unlike conventional perfect QEC codes, this code does not correct all possible single-qubit errors exactly, but instead offers \emph{approximate correction} optimized for amplitude damping noise. The recovery process involves detecting which qubit experienced a damping event and applying a conditional unitary to restore the state, resulting in high-fidelity recovery despite the approximate nature. This demonstrates that relaxing the strict criteria of perfect quantum error correction can lead to more efficient codes for specific noise models.

\section{\label{sec:noise}Effects of Noise on Multiparty QSSCM Protocol}

Quantum states are fragile to noise. Therefore, studying noises and investigating their actual effect on a protocol is crucial for implementing the protocol.

\subsection{\label{ssec:noise_QSSCM}Effect of noise on $3$-party QSSCM Protocol}

There are three different channels, namely, Bob to Charlie, Charlie to Alice and Alice to Charlie, in the QSSCM protocol we are considering here. Any of these three channels may get affected by the noise. Here, we consider the bit-flip, phase-flip and amplitude damping noise.

\paragraph{\label{para:flip_noise_QSSCM}Bit-flip and Phase-flip Noise}

If a state is prepared on a computational (Hadamard respectively) basis, from the discussion in Section~\ref{ssec:noise_model}, we can easily see that the phase-flip (bit-flip respectively) noise does not affect it. Therefore, over each channel, depending on the transmitted state, there is only one effective error, either bit-flip or phase-flip. Let us assume $p^B_*, p^C_*$ and $p^A_*$ are the error probabilities for the channels $\mathcal C^B$ from Bob to Charlie, $\mathcal C^C$ from Charlie to Alice and $\mathcal C^A$ from Alice to Charlie, respectively. Here, $*$ in the suffix denotes the bit-flip or phase-flip (whichever is applicable). Therefore, a state $\rho$ becomes
\begin{equation}
\mathcal{C}(\rho)=(1-p^\mathcal{C}_*)\rho+p^\mathcal{C}_*\rho'
\end{equation}
under the channel $\mathcal{C}$, where $\rho'$ is the state obtained by a bit-flip or phase-flip (whichever is applicable) error on the state $\rho$. Note that, for any operator $U$ applied by Bob, Charlie or Alice on a prepared state $\rho$,
\begin{equation}
\label{eq:commute_ops_noise}
U\rho'=(U\rho)'
\end{equation}
holds, up to some global phase $\pm1$. Thus, if Bob prepares a state as $\rho$, after going through all three channels, the final state (up to some global phase) would be
\begin{widetext}
\allowdisplaybreaks
\begin{align}
\mathcal{C}^A(\mathcal{C}^C(\mathcal{C}^B(\rho)))=&\mathcal{C}^A(\mathcal{C}^C((1-p^B_*)\rho+p^B_*\rho'))=\mathcal{C}^A((1-p^C_*)\left[(1-p^B_*)\rho+p^B_*\rho'\right]+p^C_*\left[(1-p^B_*)\rho'+p^B_*(\rho')'\right])\notag\\
=&\mathcal{C}^A((1-p^C_*)\left[(1-p^B_*)\rho+p^B_*\rho'\right]+p^C_*\left[(1-p^B_*)\rho'+p^B_*\rho\right])\notag\\
=&\mathcal{C}^A(\left[(1-p^C_*)(1-p^B_*)+p^C_*p^B_*\right]\rho+\left[(1-p^C_*)p^B_*+p^C_*(1-p^B_*)\right]\rho')\notag\\
=&\left[(1-p^A_*)(1-p^C_*)(1-p^B_*)+(1-p^A_*)p^C_*p^B_*+p^A_*(1-p^C_*)p^B_*+p^A_*p^C_*(1-p^B_*)\right]\rho\notag\\
&+\left[(1-p^A_*)(1-p^C_*)p^B_*+(1-p^A_*)p^C_*(1-p^B_*)+p^A_*(1-p^C_*)(1-p^B_*)+p^A_*p^C_*p^B_*\right]\rho'.\label{eq:noise_channel}
\end{align}
\end{widetext}
Here, the operations by Bob, Charlie and Alice have been ignored due to~\eqref{eq:commute_ops_noise}, we can think that all the noise acts before the unitary operations. Note that~\eqref{eq:noise_channel} is symmetric for $p^A_*, p_*^B$ and $p_*^C$. This implies that all three channels $\mathcal{C}^A, \mathcal{C}^B$ and $\mathcal{C}^C$ act similarly under bit-flip and phase-flip noise.

For simplicity, let us assume $p^A_*=p^C_*=p^B_*=p$. Then, from~\eqref{eq:noise_channel}, the probability of error for a single state will be
\begin{equation}
\label{eq:single_error}
e_1 = 3p(1-p)^2+p^3=3p(1-2p)+\mathcal{O}(p^3).
\end{equation}

\paragraph{\label{para:amp_damp_QSSCM}Effect of Amplitude Damping Noise}

We can see that the amplitude damping channel $\mathcal{C}_a$ does not commute or anticommute with the operations $U_A, U_B$ and $U_C$. In this case, the final state would be
\begin{equation}
\label{eq:damp_final}
U_B^\dagger U_C^\dagger\mathcal{C}^A_a\left(U_A\mathcal{C}^C_a\left(U_C\mathcal{C}^B_a\left(U_B\ketbra{0}U_B^\dagger\right)U_C^\dagger\right)U_A^\dagger\right)U_CU_B.
\end{equation}
There are $4$ choices for $U_B$, $3$ choices for $U_C$ and $2$ choices for $U_A$, producing $4*3*2=24$ different final states, each with probability $\frac{1}{24}$. Note that if the channels are considered as noise-free, these states would be either $\ketbra{0}$ or $\sigma_y\ketbra{0}\sigma_y=\ketbra{1}$.


\begin{table*}
\caption{Errors in the final output generated by Bob and Charlie, under amplitude damping noise with damping strength $\gamma$, for different operations applied by Alice, Bob and Charlie. Each of these cases may appear with probability $1/24$.}
\label{tab:damp_err}
\begin{center}
\begin{tabular}{|c||c||c||c|}
\hline
\textbf{Bob's State}&\textbf{Charlie's Operation}&\textbf{Alice's Secret}&\textbf{Error Probability}\\
\hline
\multirow{6}{*}{$\ketbra{0}$}&\multirow{2}{*}{$I$}&$0$&$0$\\
\cline{3-4}
&&$1$&$\gamma$\\
\cline{2-4}
&\multirow{2}{*}{$\sigma_y$}&$0$&$2\gamma-\gamma^2$\\
\cline{3-4}
&&$1$&$\gamma-\gamma^2$\\
\cline{2-4}
&\multirow{2}{*}{$H$}&$0$&$\gamma/2$\\
\cline{3-4}
&&$1$&$\gamma/2$\\
\hline
\multirow{6}{*}{$\ketbra{1}$}&\multirow{2}{*}{$I$}&$0$&$3\gamma-3\gamma^2+\gamma^3$\\
\cline{3-4}
&&$1$&$2\gamma-3\gamma^2+\gamma^3$\\
\cline{2-4}
&\multirow{2}{*}{$\sigma_y$}&$0$&$\gamma-2\gamma^2+\gamma^3$\\
\cline{3-4}
&&$1$&$2\gamma-2\gamma^2+\gamma^3$\\
\cline{2-4}
&\multirow{2}{*}{$H$}&$0$&$(3\gamma-2\gamma^2)/2$\\
\cline{3-4}
&&$1$&$(3\gamma-2\gamma^2)/2$\\
\hline
\multirow{6}{*}{$\ketbra{+}$}&\multirow{2}{*}{$I$}&$0$&$\left(1-(1-\gamma)^{3/2}\right)/2$\\
\cline{3-4}
&&$1$&$\left(1-(1-\gamma)^{3/2}\right)/2$\\
\cline{2-4}
&\multirow{2}{*}{$\sigma_y$}&$0$&$\left(1-(1-\gamma)^{3/2}\right)/2$\\
\cline{3-4}
&&$1$&$\left(1-(1-\gamma)^{3/2}\right)/2$\\
\cline{2-4}
&\multirow{2}{*}{$H$}&$0$&$\left(1-2\gamma+\gamma^2-(1-\gamma)^{5/2}\right)/2$\\
\cline{3-4}
&&$1$&$\left(1+\gamma^2-(1-\gamma)^{5/2}\right)/2$\\
\hline
\multirow{6}{*}{$\ketbra{-}$}&\multirow{2}{*}{$I$}&$0$&$\left(1-(1-\gamma)^{3/2}\right)/2$\\
\cline{3-4}
&&$1$&$\left(1-(1-\gamma)^{3/2}\right)/2$\\
\cline{2-4}
&\multirow{2}{*}{$\sigma_y$}&$0$&$\left(1-(1-\gamma)^{3/2}\right)/2$\\
\cline{3-4}
&&$1$&$\left(1-(1-\gamma)^{3/2}\right)/2$\\
\cline{2-4}
&\multirow{2}{*}{$H$}&$0$&$\left(1+2\gamma-\gamma^2-(1-\gamma)^{5/2}\right)/2$\\
\cline{3-4}
&&$1$&$\left(1-\gamma^2-(1-\gamma)^{5/2}\right)/2$\\
\hline
\multicolumn{3}{|c||}{\textbf{Average Error}}&$\left(3+8\gamma-7\gamma^2+2\gamma^3-2(1-\gamma)^{3/2}-(1-\gamma)^{5/2}\right)/12$\\
\hline
\end{tabular}
\end{center}
\end{table*}

The probability of average error under the amplitude damping channel is thus given by
\begin{align}
e_1^a&=\frac{1}{12}\left(3+8\gamma-7\gamma^2+2\gamma^3-2(1-\gamma)^{3/2}-(1-\gamma)^{5/2}\right)\notag\\
&=\frac{27}{24}\gamma-\frac{77}{96}\gamma^2+\mathcal{O}(\gamma^3).\label{eq:single_damp}
\end{align}
The probabilities of errors for all $24$ cases are explicitly mentioned in Table~\ref{tab:damp_err}.

\begin{figure}[thpb]
\centering
\includegraphics[width=\columnwidth]{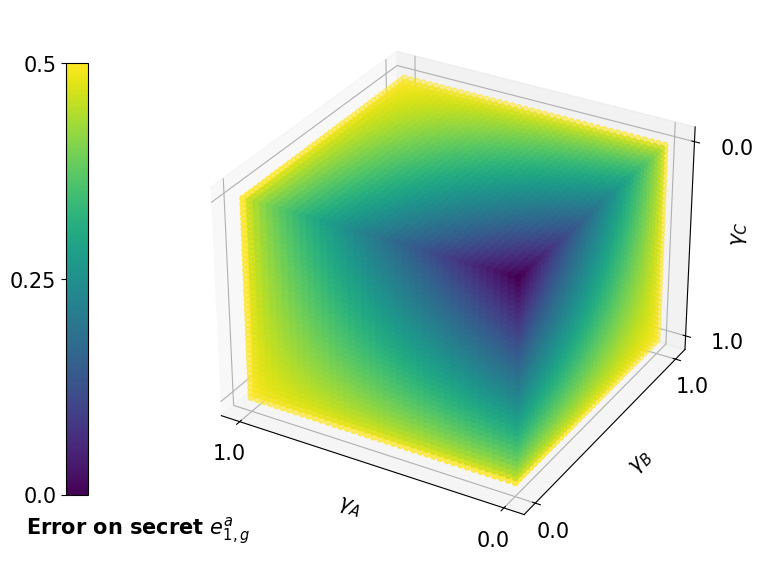}
\caption{Error on secret $e^a_{1,g}$ as function~\eqref{eq:gen_damp} of damping strengths $\gamma_A, \gamma_B$ and $\gamma_C$ for channels from Alice to Charlie, from Bob to Charlie and from Charlie to Alice, respectively. Observe that $\gamma_A$ and $\gamma_C$ affect similarly, while $\gamma_B$ effects differently.}
\label{fig:err_amp_gen}
\end{figure}

For a more general case, when the damping strengths of the channels $\mathcal{C}^A, \mathcal{C}^B$ and $\mathcal{C}^C$ are given by $\gamma_A, \gamma_B$ and $\gamma_C$, respectively, the probabilities of errors are given by  
\begin{align}
e_{1,g}^a=&\Big(4+2(\gamma_A+\gamma_B+\gamma_C)-2(\gamma_A\gamma_B+\gamma_B\gamma_C+\gamma_C\gamma_A)\notag\\
&+2\gamma_A\gamma_B\gamma_C-(1-\gamma_A)(1-\gamma_C)\sqrt{1-\gamma_B}\notag\\
&-(1-\gamma_B)\sqrt{(1-\gamma_A)(1-\gamma_C)}\notag\\
&-2\sqrt{(1-\gamma_A)(1-\gamma_B)(1-\gamma_C)}\Big)/12.\label{eq:gen_damp}
\end{align}
The probabilities of errors for different cases can be found in Table~\ref{tab:damp_err_gen}. From~\eqref{eq:gen_damp}, we can see that the effects of $\gamma_A$ and $\gamma_C$ are the same. However, $\gamma_B$ creates more error than $\gamma_A$ and $\gamma_C$. The effect of $\gamma_A, \gamma_B$ and $\gamma_C$ can be seen in Fig.~\ref{fig:err_amp_gen}.

\begin{table*}
\caption{Errors in the final output generated by Bob and Charlie, under amplitude damping noise with damping strengths $\gamma_A,\gamma_B$ and $\gamma_C$ corresponding to the channels from Alice to Charlie, from Bob to Charlie and from Charlie to Alice, respectively, for different operations applied by Alice, Bob and Charlie. Each of these cases may appear with probability $1/24$.}
\label{tab:damp_err_gen}
\begin{center}
\begin{tabular}{|c||c||c||c|}
\hline
\textbf{Bob's State}&\textbf{Charlie's Operation}&\textbf{Alice's Secret}&\textbf{Error Probability}\\
\hline
\multirow{6}{*}{$\ketbra{0}$}&\multirow{2}{*}{$I$}&$0$&$0$\\
\cline{3-4}
&&$1$&$\gamma_A$\\
\cline{2-4}
&\multirow{2}{*}{$\sigma_y$}&$0$&$\gamma_A+\gamma_C-\gamma_A\gamma_C$\\
\cline{3-4}
&&$1$&$\gamma_C-\gamma_A\gamma_C$\\
\cline{2-4}
&\multirow{2}{*}{$H$}&$0$&$\left(1-\sqrt{(1-\gamma_A)(1-\gamma_C)}\right)/2$\\
\cline{3-4}
&&$1$&$\left(1-\sqrt{(1-\gamma_A)(1-\gamma_C)}\right)/2$\\
\hline
\multirow{6}{*}{$\ketbra{1}$}&\multirow{2}{*}{$I$}&$0$&$\gamma_A+\gamma_B+\gamma_C-\gamma_A\gamma_B-\gamma_B\gamma_C-\gamma_C\gamma_A+\gamma_A\gamma_B\gamma_C$\\
\cline{3-4}
&&$1$&$\gamma_B+\gamma_C-\gamma_A\gamma_B-\gamma_B\gamma_C-\gamma_C\gamma_A+\gamma_A\gamma_B\gamma_C$\\
\cline{2-4}
&\multirow{2}{*}{$\sigma_y$}&$0$&$\gamma_B-\gamma_A\gamma_B-\gamma_B\gamma_C+\gamma_A\gamma_B\gamma_C$\\
\cline{3-4}
&&$1$&$\gamma_A+\gamma_B-\gamma_A\gamma_B-\gamma_B\gamma_C+\gamma_A\gamma_B\gamma_C$\\
\cline{2-4}
&\multirow{2}{*}{$H$}&$0$&$\left(1+(2\gamma_B-1)\sqrt{(1-\gamma_A)(1-\gamma_C)}\right)/2$\\
\cline{3-4}
&&$1$&$\left(1+(2\gamma_B-1)\sqrt{(1-\gamma_A)(1-\gamma_C)}\right)/2$\\
\hline
\multirow{6}{*}{$\ketbra{+}$}&\multirow{2}{*}{$I$}&$0$&$\left(1-\sqrt{(1-\gamma_A)(1-\gamma_B)(1-\gamma_C)}\right)/2$\\
\cline{3-4}
&&$1$&$\left(1-\sqrt{(1-\gamma_A)(1-\gamma_B)(1-\gamma_C)}\right)/2$\\
\cline{2-4}
&\multirow{2}{*}{$\sigma_y$}&$0$&$\left(1-\sqrt{(1-\gamma_A)(1-\gamma_B)(1-\gamma_C)}\right)/2$\\
\cline{3-4}
&&$1$&$\left(1-\sqrt{(1-\gamma_A)(1-\gamma_B)(1-\gamma_C)}\right)/2$\\
\cline{2-4}
&\multirow{2}{*}{$H$}&$0$&$\left(1-\gamma_A-\gamma_C+\gamma_A\gamma_C-(1-\gamma_A)(1-\gamma_C)\sqrt{1-\gamma_B}\right)/2$\\
\cline{3-4}
&&$1$&$\left(1+\gamma_A-\gamma_C+\gamma_A\gamma_C-(1-\gamma_A)(1-\gamma_C)\sqrt{1-\gamma_B}\right)/2$\\
\hline
\multirow{6}{*}{$\ketbra{-}$}&\multirow{2}{*}{$I$}&$0$&$\left(1-\sqrt{(1-\gamma_A)(1-\gamma_B)(1-\gamma_C)}\right)/2$\\
\cline{3-4}
&&$1$&$\left(1-\sqrt{(1-\gamma_A)(1-\gamma_B)(1-\gamma_C)}\right)/2$\\
\cline{2-4}
&\multirow{2}{*}{$\sigma_y$}&$0$&$\left(1-\sqrt{(1-\gamma_A)(1-\gamma_B)(1-\gamma_C)}\right)/2$\\
\cline{3-4}
&&$1$&$\left(1-\sqrt{(1-\gamma_A)(1-\gamma_B)(1-\gamma_C)}\right)/2$\\
\cline{2-4}
&\multirow{2}{*}{$H$}&$0$&$\left(1+\gamma_A+\gamma_C-\gamma_A\gamma_C-(1-\gamma_A)(1-\gamma_C)\sqrt{1-\gamma_B}\right)/2$\\
\cline{3-4}
&&$1$&$\left(1-\gamma_A+\gamma_C-\gamma_A\gamma_C-(1-\gamma_A)(1-\gamma_C)\sqrt{1-\gamma_B}\right)/2$\\
\hline
\multicolumn{3}{|c||}{\multirow{3}{*}{\textbf{Average Error}}}&$\Big(4+2(\gamma_A+\gamma_B+\gamma_C)-2(\gamma_A\gamma_B+\gamma_B\gamma_C+\gamma_C\gamma_A)+2\gamma_A\gamma_B\gamma_C$\\
\multicolumn{3}{|c||}{}&$-(1-\gamma_A)(1-\gamma_C)\sqrt{1-\gamma_B}-(1-\gamma_B)\sqrt{(1-\gamma_A)(1-\gamma_C)}$\\
\multicolumn{3}{|c||}{}&$-2\sqrt{(1-\gamma_A)(1-\gamma_B)(1-\gamma_C)}\Big)/12$\\
\hline
\end{tabular}
\end{center}
\end{table*}

\subsection{\label{ssec:noise_n_QSSCM}Effect of noise on $n$-party QSSCM Protocol}

There are $n$ different channels in $n$-party QSSCM protocol. As we have already seen, all of these $n$ channels act similarly under bit-flip and phase-flip noise; for simplicity, let us consider that they all have the same error probability $p$. Then generalizing~\eqref{eq:noise_channel} and~\eqref{eq:single_error}, we get the error for a single qubit as
\begin{align}
e_1^g&=\sum_{i\text{ is odd }\leq n}{n\choose i}p^i(1-p)^{n-i}\notag\\
&=\frac{1}{2}\left((1-p+p)^n-(1-p-p)^n\right)\notag\\
&=\frac{1}{2}\left(1-(1-2p)^n\right).\label{eq:gen_flip_error}
\end{align}

\begin{figure}[thpb]
\centering
\includegraphics[width=\columnwidth]{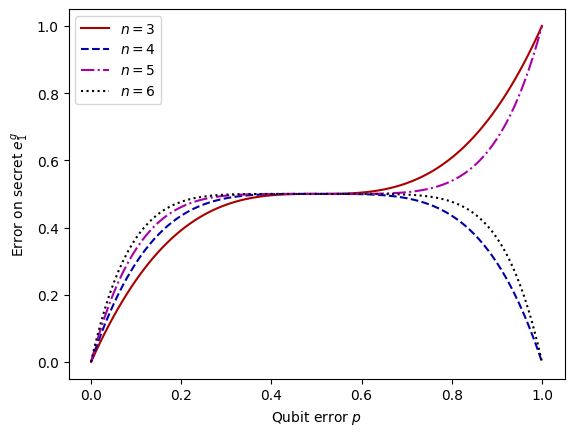}
\caption{The plots of the error $e_1^g$ as a function~\eqref{eq:gen_flip_error} of the qubit error probability $p$ are shown for $n = 3, 4, 5,$ and $6$. For protocols involving an even number of channels, a higher error probability increases the likelihood of an even number of flips, which can paradoxically lead to a reduction in the overall error. However, this does not happen for odd number of channels, leading the error on secret to $1$, as qubit error probability reaches to $1$.}
\label{fig:err_flip_gen}
\end{figure}

Fig.~\ref{fig:err_flip_gen} shows the plots of the error~$e_1^g$ against error probability~$p$ for $n=3, 4, 5$ and $6$. Note that even number of flips result in \emph{no error}. Therefore, if the error probability is high, the probability of even number of flips is also high for even numbers of channels, reducing the overall error.

\section{\label{sec:QEC}Improvement of Result Using QEC}

Quantum error correction (QEC) is a crucial technique in quantum computing designed to protect quantum information from noise and errors that arise due to imperfections in quantum systems. Unlike classical error correction, where bits are used to represent information, quantum error correction must account for both bit-flip and phase-flip errors, as well as more complex quantum errors that affect quantum states, such as coherence.


\subsection{\label{ssec:Shor_rep}Shor's Code as Repetition Code}

Shor's code is not only very easy to implement but it is also efficient for the QSSCM code we are considering here. As we already discussed in Section~\ref{ssec:noise_QSSCM}, we do not require to combine the bit-flip and phase-flip correction together, rather we apply bit-flip correction for the states $\ket{0}$ and $\ket{1}$ and phase-flip correction for the states $\ket{+}$ and $\ket{-}$. Therefore, we only require to repeat each state three times, reducing the resource requirement for Shor's code from 9 to 3. This also makes the code resource efficient compared to other QEC codes, where we cannot separate bit-flip and phase-flip correction, requiring at least $5$ physical qubits due to the quantum singleton bound~\cite{DJORDJEVIC2012227}. As our modified Shor's code is only repeating the states, we would call it \emph{repetition code}. Under this QEC scenario, Alice, Bob and Charie randomly choose one operation as described in the QSSCM protocol and apply this operation on three consecutive qubits. During decoding, the secret is decided based on majority voting among three consecutive states. Thus the components of the repetition code is as follow.

\textbf{Encoding: }Prepare three copies of each state during state preparation.

\textbf{State Recovery: }Measure each state in proper basis, as prepared. Apply majority voting to decide the measurement outcome.

\textbf{Logical Operation: }Apply each operation on three consecutive states.

 This repetition code can also be used for other single-qubit-based quantum protocols~\cite{GUO2003247,refId0,PhysRevA.92.030301,1573668924803137792,PhysRevLett.68.3121, PhysRevA.76.062316, PhysRevLett.92.057901,PhysRevA.69.052319,WANG2006256,10.1007/s11128-017-1757-x,JiXin1418, 10.1007/s11128-014-0767-1,4099193,doi:10.1142/S0219749916500027,PhysRevA.62.022305,10.1007/s11128-018-2124-2}, where the outcome of the protocol is a sequence of classical bits, to improve the result.

\subsection{\label{ssec:QSSCM_rep}Repetition Code on QSSCM Protocol}

If at most one from the measurements of three consecutive states gives erroneous output, the majority voting decoder would provide the correct secret. However, if more than one output is erroneous, this decoder would provide the wrong secret bit. If $e$ is the error for a single qubit, the error after QEC would be given by
\begin{equation}
\label{eq:QEC_final_error}
e_{QEC}=3e^2(1-e)+e^3.
\end{equation}
The error correction would be effective, if $e>0$ and $e_{QEC}<e$, that is,
\begin{align}
3e^2(1-e)+e^3<e&\implies3e(1-e)+e^2<1\notag\\
&\implies2e^2-3e+1>0\notag\\
&\implies\left(e-\frac{1}{2}\right)(e-1)>0\notag\\
&\implies e<\frac{1}{2},\;\;\text{ as }e\leq1.\label{eq:QEC_cond}
\end{align}
This is the condition for effective error correction using the repetition code.

\paragraph{\label{para:QEC_flip_noise}Bit-flip and Phase-flip Noise}

\begin{figure*}[thpb]
\centering
\includegraphics[width=0.95\textwidth]{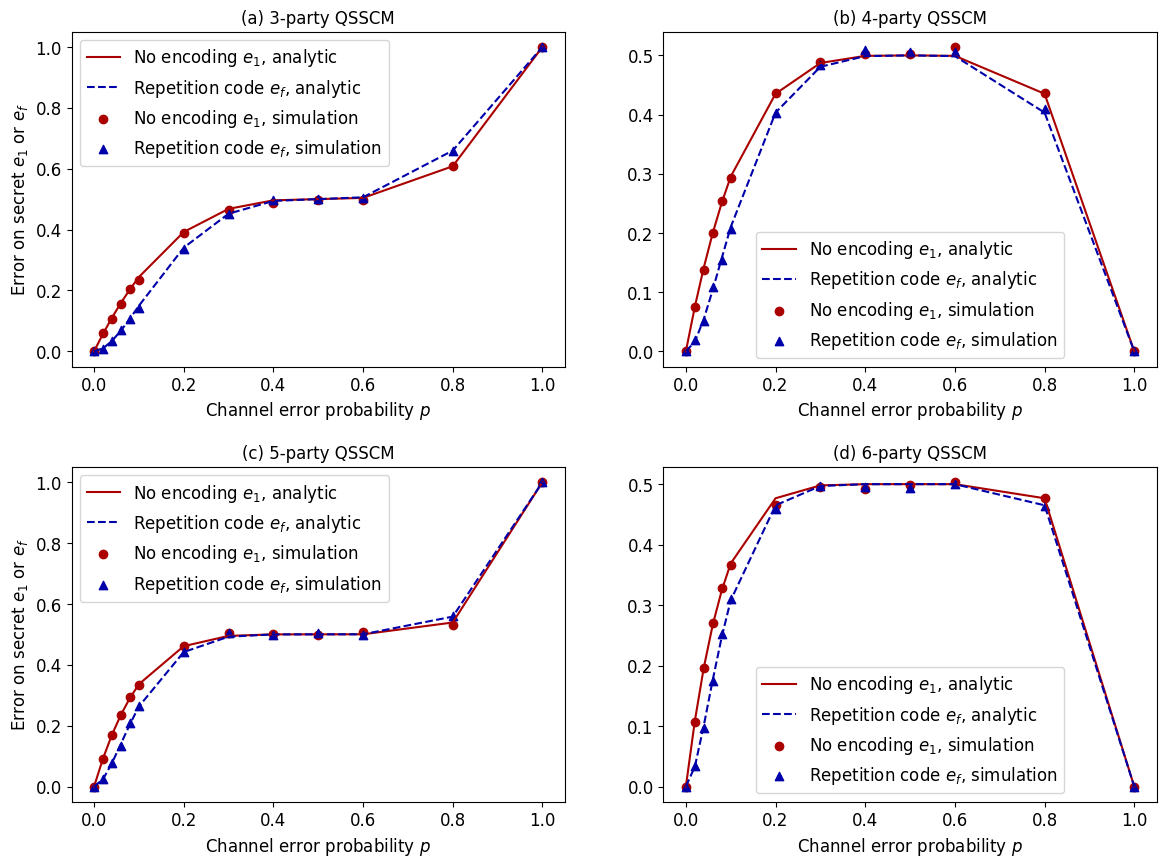}
\caption{Error on reconstructed secret is plotted against channel error probability $p$ with and without repetition code. The plot is generated from analytic equations~\eqref{eq:gen_flip_error} and~\eqref{eq:QEC_gen_error} and the simulated results. It shows that if $p<0.5$, the repetition code can reduce the error in the reconstructed secret.}
\label{fig:rep_flip}
\end{figure*}

As we have discussed above, if more than one state from three consecutive states gets flipped, the state cannot be restored perfectly, leading to an error. Now, for $3$-party QSSCM, using~\eqref{eq:single_error}, we can write the  probability that at least two out of three consecutive states get flipped as
\begin{align}
e_f&=3e_1^2(1-e_1)+e_1^3=27p^2+\mathcal{O}(p^3),\label{eq:QEC_error}
\end{align}
which is the final error probability using the repetition code.

From~\eqref{eq:QEC_cond}, the condition for an effective error correction under bit-flip and phase-flip noise is given by
\begin{align}
e_1<\frac{1}{2}&\implies3p(1-p)^2+p^3<\frac{1}{2}\notag\\
&\implies\frac{1}{2}\left(1-(1-2p)^3\right)<\frac{1}{2}\notag\\
&\implies p<\frac{1}{2}.
\end{align}
Therefore, if all three channels have an error probability less than $\frac{1}{2}$, the repetition code can effectively reduce the error, which is shown in Fig.~\ref{fig:rep_flip}(a). The result after simulating QSSCM under bit-flip and phase-flip noise is also shown in the same figure. The simulation plot also shows that if $p<0.5$, the repetition code can reduce the error.

For $n$-party QSSCM protocol, using~\eqref{eq:gen_flip_error}, we can write the error on secret as
\begin{align}
e^g_f&=3(e^g_1)^2(1-e^g_1)+(e^g_1)^3\notag\\
&=\frac{1}{4}\left(1-(1-2p)^n\right)^2\left(2+(1-2p)^n\right).\label{eq:QEC_gen_error}
\end{align}
Also, the condition for effective error correction becomes
\begin{align}
e_1^g<\frac{1}{2}&\implies\frac{1}{2}\left(1-(1-2p)^n\right)<\frac{1}{2}\notag\\
&\implies(1-2p)^n>0\notag\\
&\implies\begin{cases}
p\in(0,1)\backslash{\frac{1}{2}},\;&\text{if $n$ is even},\\
p\in(0,\frac{1}{2})&\text{if $n$ is odd}.
\end{cases}
\end{align}
This result along with the simulations for $n=3, 4, 5$ and $6$ has been shown in Fig.~\ref{fig:rep_flip}.

\paragraph{\label{para:QEC_damp_noise}Amplitude Damping Noise}

\begin{figure*}[thpb]
\centering
\includegraphics[width=0.95\textwidth]{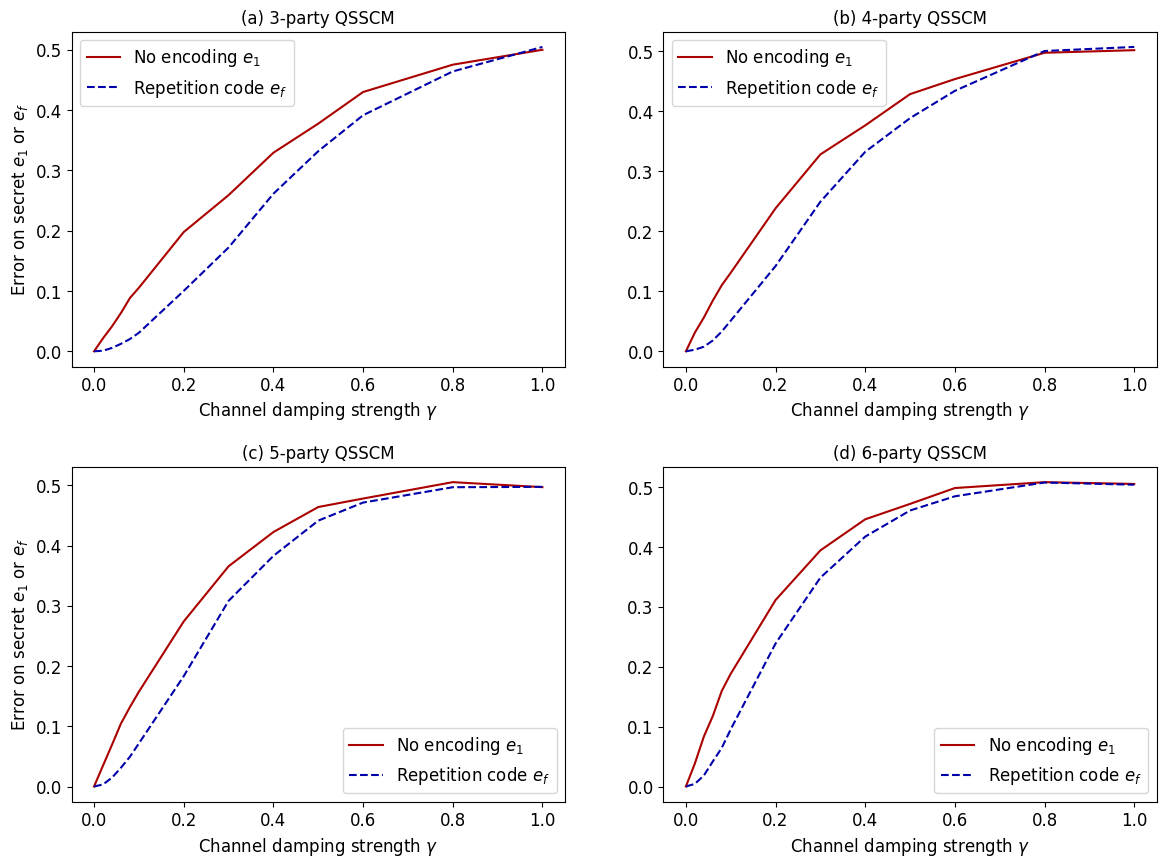}
\caption{Error on reconstructed secret is plotted against damping probability $\gamma$ with and without repetition code. The plot is generated from analytic equations~\eqref{eq:single_damp} and~\eqref{eq:QEC_damp} and the simulated results. It shows that the repetition code can reduce the error in the reconstructed secret for all values of the damping strength.}
\label{fig:rep_amp_gen}
\end{figure*}

For amplitude damping noise, if only one from three consecutive outputs is erroneous, the majority voting decoder would provide the correct secret bit. However, if two or three outputs are erroneous, then the majority voting would fail. Using~\eqref{eq:single_damp}, the probability of at least two out of three consecutive outputs being erroneous can be written as
\begin{align}
e_f^a&=3\left(e_1^a\right)^2(1-e_1^a)+\left(e_1^a\right)^3=\frac{3}{32}\gamma^2+\mathcal{O}(\gamma^3).\label{eq:QEC_damp}
\end{align}

From~\eqref{eq:single_damp} one can see that, $e_1^a$ satisfies the condition for effective error correction~\eqref{eq:QEC_cond} for $\gamma\in(0, 1)$, that is,
\begin{equation}
e_1^a<\frac{1}{2}\;\text{for}\;\gamma\in(0, 1).
\end{equation}

Also, $e_{1,g}^a$ in~\eqref{eq:gen_damp} satisfies the effective error correcting condition for all $\gamma_A, \gamma_B, \gamma_C\in(0, 1)$. We have simulated the $3$-party QSSCM protocol with amplitude damping noise. We see that the repetition code reduces the error in the reconstructed secret for all values of damping strength except $0$ and $1$, where it is the same as the error for the no-encoding scenario.

We have also simulated the $n$-party QSSCM protocol for $n=3, 4, 5$ and $6$ under the amplitude damping noise. The plots are shown in Fig.~\ref{fig:rep_amp_gen}.

\subsection{\label{ssec:QSSCM_QEC45}Five and Four-qubit QECs on QSSCM Protocol}

\begin{figure}[thpb]
\centering
\includegraphics[width=\columnwidth]{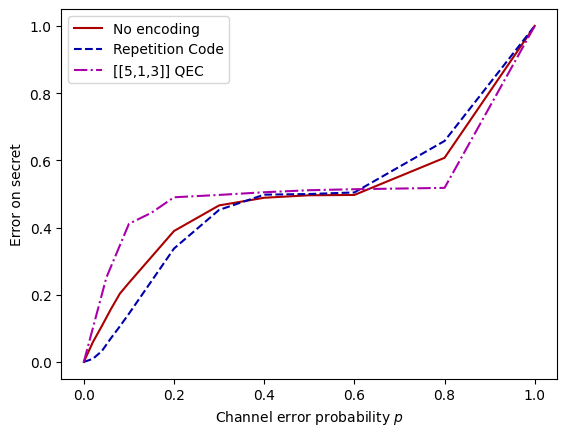}
\caption{Plots show the simulated errors on reconstructed secret against Pauli (bit-flip or phase-flip) error $p$. Observe that repetition code performs better than existing $\llbracket5,1,3\rrbracket$ perfect QEC~\cite{laflamme1996perfect}. Even the five-qubit codes perform worse than the no encoding scenario. The reason is for five-qubit code, all the five qubits are going through the error, making the error very high compared to one-qubit error in no encoding scenario, and the QEC fails to recover it. Although for the repetition code, three qubits are going through the error, the QEC restricts it below the no encoding threshold.}
\label{fig:QEC_flip}
\end{figure}

\begin{figure}[thpb]
\centering
\includegraphics[width=\columnwidth]{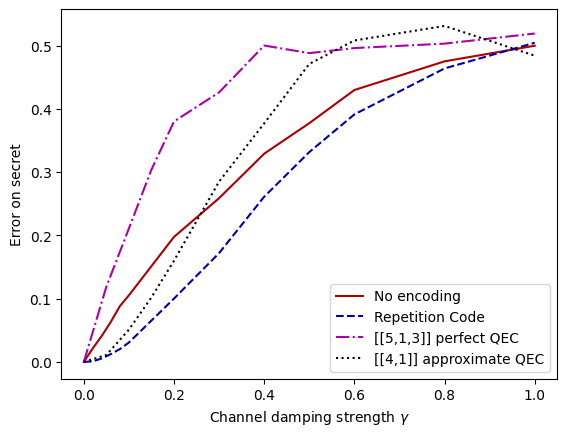}
\caption{Plots show the simulated error of the secret as a function of the amplitude damping strength, $\gamma$. Notably, the repetition code outperforms both the $\llbracket5,1,3\rrbracket$ perfect QEC code~\cite{laflamme1996perfect} and the $\llbracket4,1\rrbracket$ approximate QEC code~\cite{leung1997approximate}. Both the four-qubit and five-qubit codes perform worse than the no-encoding scenario. This is because, during quantum error correction, all qubits in these codes are exposed to noise, resulting in a higher cumulative error than the single-qubit error encountered without encoding. In contrast, although the repetition code also subjects three qubits to noise, the QEC process effectively suppresses the total error below that of the no-encoding case.}
\label{fig:QEC_amp}
\end{figure}

In the previous section, we have applied encoding at the beginning of the protocol, and the decoding including recovery operation at the end of the complete protocol. As there is a single cycle of encoding, decoding and state recovery operations, we call this as \emph{single-cycle QEC}. However, we can apply this QEC in multiple cycles, which performs better than a single-cycle QEC~\cite{PhysRevResearch.5.043161,doi:10.1126/science.1203329,Cramer2016,Kelly2015,Andersen2020,Krinner2022}. One straightforward towards multiple cycles is to apply a single cycle to each channel individually, first from Bob to Charlie, then from Charlie to Dave, and so on. In this scenario, each party can apply their physical operations directly on the main qubits after the recovery operations. Note that we cannot apply multiple cycles for the repetition code we discussed in Section~\ref{ssec:Shor_rep} as that requires the knowledge about the basis on which Bob prepares the state. However, the five-qubit perfect code~\cite{laflamme1996perfect} and the four-qubit approximate code~\cite{leung1997approximate} being basis independent, this problem does not arise, and we can perform the multiple cycles. Fig.~\ref{fig:QEC_flip} and~\ref{fig:QEC_amp} show the performances of the five-qubit perfect code and the four-qubit approximate code in multi-cycle scenario for $3$-party QSSCM under Pauli (bit-flip and phase-flip) noise and amplitude damping noise, respectively. Even the four-qubit and the five-qubit code performs worse than the no-encoding scenario. This is because, in the five-qubit code, all five qubits are subjected to errors, resulting in a higher overall error rate compared to the single-qubit error in the no-encoding case. In contrast, while the repetition code also exposes three qubits to errors, QEC keeps the total error rate below the threshold of the no-encoding scenario.

\section{\label{sec:SSQI}SSQI Protocol under Noise}

Zhang et al. also proposed a SSQI protocol in the same article~\cite{PhysRevA.71.044301} combining the above QSSCM protocol with the standard teleportation protocol~\cite{PhysRevLett.70.1895}. To perform the secret sharing of a quantum state among $n-1$ receivers, Alice sends the state to Bob using the standard teleportation protocol. However, instead of announcing the Bell-measurement outcomes, she shares these among the other $n-2$ receivers except Bob using the QSSCM protocol. Therefore, this SSQI protocol requires QEC for the QSSCM part as well as for the standard teleportation. Several fault-tolerant teleportation schemes~\cite{stack2025assessingteleportationlogicalqubits,ataides2025constantoverheadfaulttolerantbellpairdistillation,PhysRevLett.76.722,doi:10.1126/science.adp6016,Bouwmeester1997,shalby2025optimizednoiseresilientsurfacecode} have been proposed to deal with the noise during quantum teleportation. However, if $e_t$ is the error in the teleportation process, and $e_{noise}$ and $e_{QEC}$ are the errors for the QSSCM protocol, without and with QEC, respectively, for effective error correction we require
\begin{align}
&\text{fidelity with correction}>\text{fidelity without correction}\notag\\
&\implies(1-e_{QEC})(1-e_t)>(1-e_{noise})(1-e_t)\notag\\
&\implies e_{QEC}<e_{noise},
\end{align}
which is the effective error correcting condition for the QSSCM protocol. Therefore, all the results we have discussed in Section~\ref{sec:QEC} are also valid for the SSQI protocol.

\section{\label{sec:conclusion}CONCLUSION AND FUTURE WORKS}

In this article, we investigate the effects of quantum noise on the multiparty QSSCM and SSQI protocols proposed by Zhang et al.~\cite{PhysRevA.71.044301}. The QSSCM protocol utilizes single-qubit transmissions without the need for entanglement, offering a relatively simple and practical implementation. The SSQI protocol builds upon QSSCM to enable the sharing of quantum information. Despite their simplicity, these protocols are highly vulnerable to quantum noise, which can corrupt the transmitted qubits and significantly hinder the accurate reconstruction of the secret.

To address the vulnerability of the QSSCM protocol to quantum noise, we analyze the impact of various noise models—including bit-flip, phase-flip, and amplitude damping—on its performance. Our analysis demonstrates how these noise sources introduce errors that degrade the fidelity of the reconstructed secret. To mitigate these effects, we apply an optimized version of Shor’s 9-qubit quantum error correction (QEC) code. By separating the bit-flip and phase-flip correction processes, we reduce the required resources from 9 qubits to just 3. This simplified, repetition-based QEC approach significantly lowers the error probability compared to conventional QEC schemes, thereby enhancing the robustness of the QSSCM protocol against quantum noise. In general, such a 3-qubit abridged version of Shor's code is not capable of correcting amplitude damping noise. However, in this context, it proves effective due to the specific structure of the QSS protocol. Our findings and methodology are equally applicable to the SSQI protocol, and we argue that the proposed QEC technique can be extended to other single-qubit-based quantum protocols.

Future research could investigate more efficient quantum error correction techniques—such as surface codes or optimized encoding strategies—to further improve the security and practicality of multiparty quantum secret sharing in realistic, noisy environments. Additionally, the application of repetition-based error correction could be extended to other single-qubit-based quantum protocols, including QSS, QKD, QSDC, and QA schemes, to evaluate its effectiveness in enhancing their resilience to noise.



%


\end{document}